\documentclass[11pt]{iopart}
\usepackage[square,sort&compress,numbers]{natbib}
\usepackage{iopams}  
\usepackage{pstricks-add}
\usepackage[colorlinks=true,citecolor=blue]{hyperref}
\usepackage{cleveref}
\usepackage{setstack}
\usepackage{etoolbox}
\usepackage[perpage]{footmisc}
\usepackage{graphicx}

\crefname{equation}{}{}
\patchcmd{\numparts}{\addtocounter{equation}{1}}{\refstepcounter{equation}}{}{}

\newcommand*{\diff}{{\mathrm d}}

\makeatletter
\renewcommand{\dddot}[1]{%
  {\mathop{\kern\z@#1}\limits^{\vbox to-1.4\ex@{\kern-\tw@\ex@
   \hbox{\normalfont ...}\vss}}}}
\makeatother

\begin{document}

\title[On the interplay between Noether's theorem and the theory of adiabatic invariants]{On the interplay between Noether's theorem and the theory of adiabatic invariants}

\author{Thierry Gourieux and Rapha\"el Leone}

\address{Laboratoire de Physique et Chimie Th\'eoriques (UMR CNRS 7019), Universit\'e de Lorraine, F-54506 Vand\oe uvre-l\`es-Nancy}

\ead{\href{mailto:thierry.gourieux@univ-lorraine.fr}{thierry.gourieux@univ-lorraine.fr}}

\begin{abstract}
This article focuses on an important quantity that will be called the Rund-Trautman function. It already plays a central role in Noether's theorem since its vanishing characterizes a symmetry and leads to a conservation law. The main aim of the paper is to show how, in the realm of classical mechanics, an `almost' vanishing Rund-Trautman function accompanying an `almost' symmetry leads to an `almost' constant of motion within the adiabatic assumption, that is, to an adiabatic invariant. To this end, the Rund-Trautman function is first introduced and analysed in detail, then it is implemented for the general one-dimensional problem. Finally, its relevance in the adiabatic context is examined through the example of the harmonic oscillator with a slowly varying frequency. Notably, for some frequency profiles, explicit expansions of adiabatic invariants are derived through it and an illustrative numerical test is realized.  
\end{abstract}

%
\vspace{2pt}

\section{Introduction}

It is an understatement to say that, for a century now, Noether's two theorems \cite{Noether,Bessel-Hagen} have played a central role in physics \cite{YKS}. Since both symmetry and variational principles have become so fundamental to the point that they structure modern theories, these theorems are nowadays an indispensable part of any syllabus in pure physics. At an advanced level: in gauge theories of which they constitute a cornerstone, but also, beforehand, in analytical mechanics where Noether's name is generally mentioned for the first time through her first theorem. Indeed, the second theorem objectively plays a marginal role in this framework, though it is not without interest concerning parameter-free variational principles \cite{Logan} (as Jacobi's formulation of the principle of least action). The first theorem, for its part, is simply called `the theorem of Noether', and is mostly stated with the principal aim of rederiving the usual conservation laws from invariances of the action under transformations. It has the advantage of gathering all the existing conservation laws under a universal and elegant symmetry principle.

A semester-length course on classical mechanics --- which must cover many topics such as calculus of variations, Lagrangian and Hamiltonian mechanics, canonical transformations, the Hamilton-Jacobi equation, action-angle coordinates, and so on, not to mention all the important applications --- cannot spend many time on each individual chapter. This is the reason why the first theorem of Noether is rarely treated in more detail. Pedagogical articles on the subject are aimed to fill this gap. One of the most interesting topics is certainly the reversed question about searching for unknown conservation laws in some practical situation, by analysing the symmetries (if any) admitted by the action functional. The method, as described for example in Neuenschwander's book \cite{Neuenschwander}, consists in seeking transformations satisfying the so-called Rund-Trautman identity \cite{Rund,Trautman}. That task can generally be achieved in an algorithmic way if we restrict ourselves to point transformations \cite{Lutzky,Prince,Gourieux}. 

It is interesting to interpret the Rund-Trautman identity as the identical vanishing of a quantity $R$ that will be called the \emph{Rund-Trautman function} in this article. As will be shown, $R$ turns out to be the rate of change along the motion of both (i) a Noether asymmetricity and (ii) the non conservation of the quantity that would have been conserved in case of symmetry. These properties prove the role that can play $R$ in the problem of perturbatively finding  `almost' conservation laws when a system depends on small parameters. This is the main purpose of the present article regarding the adiabatic hypothesis. Our approach will be elementary enough to avoid complicated mathematical techniques \cite{Lochak}.

The paper is organized as follows. We first draw in section \ref{sec:events} a basic portrait of the space of events associated with a mechanical system, and we outline how point transformations act in that space. In section \ref{sec:Noether} are reviewed the important features of Noether's first theorem regarding point transformations, including the Rund-Trautman function as well as coordinate and gauge issues. In section \ref{sec:applications}, we apply the theory to the determination of all the Noether point symmetries of natural one-dimensional problems \cite{Lewis-Leach}, with an emphasis put on the time-dependent harmonic oscillator. Section \ref{sec:adiabatic} is then devoted to the notions of almost symmetries and almost conservation laws within the adiabatic hypothesis. In order to comply with the amusing saying according to which `classical mechanics is the art of solving the harmonic oscillator in many ways', we treat the case of the paradigmatic time-dependent harmonic oscillator \cite{Ehrenfest,Kulsrud}. After obtaining a formal expansion of an adiabatic invariant to arbitrary orders, we derive explicit expressions for some frequency profiles and we end the article with a numerical test. 

In order to remain as pedagogical as possible without obscuring the main content, all the calculations will be explained and detailed in appendices. Alternatively, they can be considered as exercises for the reader.

\section{Continuous point transformations in the space of events}\label{sec:events}

\subsection{The space of events and its coordinatizations}\label{subsecsec:events}

Let us take as a starting point a classical mechanical problem whose configuration space $\mathcal Q$ is an $n$-dimensional smooth manifold. Since  our framework is Newtonian, there exists independently a timeline $\mathcal T$, diffeomorphic to the real line, whose points are the positions in time. The space of events (or extended configuration space) is the Cartesian product $\mathcal E=\mathcal T\times\mathcal Q$ usually coordinatized by $(1+n)$-tuples whose first element is the absolute time along $\mathcal T$, the $n$ others being coordinates in $\mathcal Q$. However, it might be interesting to consider arbitrary coordinate systems in $\mathcal E$ possibly `mixing' $\mathcal T$ and $\mathcal Q$. Such a system will be generically denoted by $(t,q)$ with $q=(q^i)$ and $i=1,\dots,n$. We must only be sure that $t$ can play the role of a time along the actual evolution of the configuration point in $\mathcal Q$, i.e.\ that it increases strictly with the absolute time (it is certainly the case if $t$ is simply a future-oriented coordinate in $\mathcal T$, e.g.\ the absolute time itself). Geometrically, it amounts to say that the curve drawn by the evolution in $\mathcal E$ is, in coordinates, the graph of a mapping $\mathcal C\colon t\mapsto q(t)$. It is often interesting to reduce at least formally the specificity of the time coordinate by setting $t\equiv q^0$ and $(t,q)=(q^\mu)$ with $\mu=0,\dots,n$.

\subsection{Continuous point transformations}

A continuous point transformation $\Phi$ of the space of events is essentially a mechanism which unambiguously maps any event $(t,q)$ into a parameter-dependent one, $(t_\varepsilon,q_\varepsilon)$ say, where $\varepsilon$ is the parameter. Locally, it admits a form 
\begin{equation}
(t,q)\longrightarrow(t_\varepsilon,q_\varepsilon)=\Phi(t,q;\varepsilon)=\big(t+\varepsilon\tau(t,q),q+\varepsilon\xi(t,q)\big)+\mathrm O(\varepsilon^2),\label{transfo}
\end{equation}
in some vicinity of $\varepsilon=0$, the quantities $\tau$ and $\xi=(\xi^i)$ being smooth functions of their arguments. The transformation $\Phi$ is entirely characterized and generated by the vector field \cite{Olver}
\begin{equation*}
\mathsf X =\tau\,\frac{\partial}{\partial t}+\xi^i\frac{\partial}{\partial q^i}=\xi^\mu\partial_\mu\qquad\Bigg(\xi^0\equiv\tau\;,\;\partial_\mu\equiv\frac{\partial}{\partial q^\mu}\Bigg)
\end{equation*}
on $\mathcal E$, the Einstein summation convention being assumed in this paper (Latin and Greek indices cover the ranges $1,\dots,n$ and $0,\dots,n$ respectively). It is the vector field which has for components the $n$-tuple $(\xi^\mu)$ with respect to the coordinate system $(q^\mu)$.

Actually, $\Phi$ drags the events along the integral curves of $\mathsf X$. From now on, $\varepsilon$ will be considered as an infinitesimal and any term of order higher than the first in $\varepsilon$ will be neglected. This way, $\varepsilon\mathsf X$ is the infinitesimal translation bringing the original event $(t,q)$
to the transformed one $(t_\varepsilon,q_\varepsilon)$. 

\subsection{Induced transformations and symmetries of point functions}

Let $F_0$ be a smooth point function\footnote{The French mathematician Gabriel Lam\'e called \textit{`fonction-de-point'} a real-valued function defined on the `absolute space' under consideration (originally the three-dimensional physical space) \cite{Lame}. It unambiguously associates a real value to any point of that space and can secondarily acquire an analytical expression through a coordinate system. It is a basic example of scalar.} defined on $\mathcal E$. Applying a first-order Taylor expansion based on \eref{transfo}, while $(t,q)$ is mapped into $(t_\varepsilon,q_\varepsilon)$, the value of $F_0$ undergoes the transformation
\begin{equation}
F_0(t,q)\longrightarrow F_0(t_\varepsilon,q_\varepsilon)=F_0(t,q)+\varepsilon\mathsf X (F_0)(t,q).\label{transpoint}
\end{equation}
One says that $\Phi$ is a symmetry of $F_0$ if it leaves its values invariant in the flow of $\Phi$, i.e.\ if the variation of $F_0$ is everywhere zero in the direction of $\mathsf X$. This is the case if and only if (iff) $\mathsf X (F_0)$ vanishes identically. Geometrically, it means that the field $\mathsf X$ is tangent to the level surfaces of $F_0$ (or equivalently that the integral curves of $\mathsf X$ are contained in these surfaces).

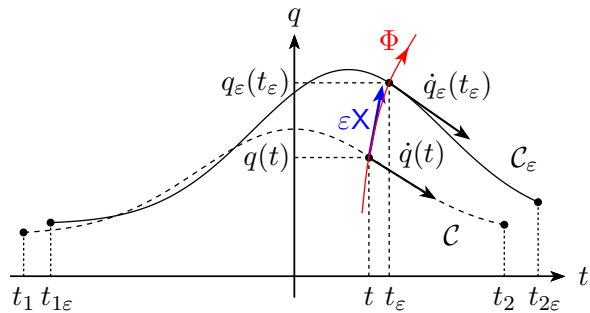
\begin{figure}
\begin{center}
\psset{xunit=1.8cm,yunit=1.4cm,algebraic=true,dimen=middle,dotstyle=o,dotsize=3pt 0,linewidth=0.5pt,arrowsize=3pt 2,arrowinset=0.25}
\begin{pspicture*}(-2.1,-0.7)(2.2,2.4)
\begin{small}
\psaxes[labelFontSize=\scriptstyle,xAxis=true,yAxis=true,labels=none,ticks=none]{->}(-0.0,-.3)(-2.1,-0.5)(2.,2.)
\psplot[plotpoints=200,linestyle=dashed,dash=2pt 2pt]{-2}{1.55}{2.718281828459045^(-x^(2.0))+0.1}
\psplot[plotpoints=200]{-1.8}{1.8}{1.2*2.718281828459045^(-(x-0.4)^(2.0)+0.2)+0.2}
\psdots[dotstyle=*](0.55,0.83)
\psline[linestyle=dashed,dash= 1.5pt 1.5pt](0.55,0.83)(.55,-.3)
\uput{0.15cm}[-90](.55,-.3){$t$}
\psline[linestyle=dashed,dash= 1.5pt 1.5pt](0.55,0.83)(0,0.83)
\uput{0.05cm}[180](0,0.83){{$q(t)$}}
\psline[linewidth=.8pt]{->}(.55,.83)(1.05,.43)
\uput{0.4cm}[105](1.05,.43){{$\dot q(t)$}}
\psdots[dotstyle=*](0.7,1.54)
\psline[linewidth=1pt,linecolor=blue]{->}(0.55,0.83)(0.65,1.52)
\psline[linestyle=dashed,dash= 1.5pt 1.5pt](0.7,1.54)(-.0,1.54)
\uput{.05cm}[180](-.0,1.54){$q_\varepsilon(t_\varepsilon)$}
\psline[linestyle=dashed,dash= 1.5pt 1.5pt](0.7,1.54)(0.7,-.3)
\uput{.15cm}[-90](0.75,-.3){$t_{\varepsilon}$}
\psline[linewidth=.8pt]{->}(0.7,1.54)(1.3,1.0)
\rput[bl](.95,1.37){$\dot q_\varepsilon(t_\varepsilon)$}
\rput[bl](1.1,0){{$\mathcal C$}}
\rput[bl](1.6,.7){{$\mathcal C_\varepsilon$}}
\rput[bl](.32,1.1){\color{blue}$\varepsilon\mathsf X$}
\uput[0](2,-.3){$t$}
\uput{.1cm}[90](0.0,2){$q$}
\psdots[dotstyle=*](-2,0.1183)
\psline[linestyle=dashed,dash= 1pt 1pt](-2,0.1183)(-2,-.3)
\uput{.15cm}[-90](-2.0,-.3){$t_1$}
\psdots[dotstyle=*](-1.8,0.212)
\psline[linestyle=dashed,dash= 1pt 1pt](-1.8,0.212)(-1.8,-.3)
\uput{.15cm}[-90](-1.75,-.3){$t_{1\varepsilon}$}
\psdots[dotstyle=*](1.8,0.406)
\psline[linestyle=dashed,dash= 1pt 1pt](1.8,0.406)(1.8,-.3)
\uput{.15cm}[-90](1.85,-.3){$t_{2\varepsilon}$}
\psdots[dotstyle=*](1.55,0.19)
\psline[linestyle=dashed,dash= 1pt 1pt](1.55,0.19)(1.55,-.3)
\uput{.15cm}[-90](1.55,-.3){$t_{2}$}
\pscurve[linecolor=red](.5,.3)(0.55,0.83)(0.7,1.54)(.9,2)
\psline[linewidth=.8pt,linecolor=red]{->}(0.840,1.87)(.841,1.8722)
\rput(.7,1.95){\color{red}$\Phi$}
\end{small}
\end{pspicture*}
\end{center} 
\caption{Under the action of $\Phi$, and for sufficiently small values of $|\varepsilon|$, the original evolution $\mathcal C$ (dashed line) is mapped into a neighbouring evolution $\mathcal C_\varepsilon$ (solid line). The vector $\varepsilon\mathsf X$ is the first order approximation in $\varepsilon$ of the transformation ($\varepsilon$ is taken positive in the figure).}\label{fig:transformation_evolution}
\end{figure}

\subsection{Induced transformations of evolutions and their kinematic properties}

Now, let $t\mapsto q(t)$ be a generic smooth evolution of the configuration between two extremities of time $t=t_1$ and $t=t_2$. As illustrated in figure \ref{fig:transformation_evolution}, for sufficiently small values of $|\varepsilon|$, its graph $\mathcal C$ drawn in $\mathcal E$ is transformed by $\Phi$ into the graph of another evolution $\mathcal C_\varepsilon$ between $t=t_{1\varepsilon}$ and $t=t_{2\varepsilon}$, according to (see \ref{app:graph})
\begin{equation}
(t,q(t))\longrightarrow (t_\varepsilon,q_\varepsilon(t_\varepsilon))=(t+\varepsilon\tau(t,q(t)),q(t)+\varepsilon\xi(t,q(t))).\label{transgraph}
\end{equation} 
Then, the velocity $\dot q(t)$ of the original evolution at $t=t$ becomes the velocity $\dot q_\varepsilon(t_\varepsilon)$ of the transformed evolution at $t=t_\varepsilon$. Viewing $t_\varepsilon$ and $q_\varepsilon(t_\varepsilon)$ as functions of the original value $t$ of the time through the equality in Equation \eref{transgraph}, one has
\begin{equation}
\dot q^i_\varepsilon(t_\varepsilon)=\frac{\diff q^i_\varepsilon(t_\varepsilon)}{\diff t_\varepsilon}=\frac{\diff q^i_\varepsilon(t_\varepsilon)}{\diff t}\Bigg\slash\frac{\diff t_\varepsilon}{\diff t}\,,\label{transvel}
\end{equation}
that is, to the first order in $\varepsilon$ (see \ref{app:vel}) :
\begin{equation}
\dot q^i_\varepsilon(t_\varepsilon)=\dot q^i(t)+\varepsilon\Bigg[\frac{\diff\xi^i(t,q(t))}{\diff t}-\dot q^i(t)\frac{\diff\tau(t,q(t))}{\diff t}\Bigg].\label{velocities}
\end{equation}
In brief, the formal transformation rules of the time, position and velocity are thus
\begin{equation*}
t\longrightarrow t_\varepsilon=t+\varepsilon \tau\,,\qquad q\longrightarrow q_\varepsilon=q+\varepsilon\xi\,,\qquad \dot q\longrightarrow \dot q_\varepsilon=\dot q+\varepsilon(\dot\xi-\dot q\dot\tau).
\end{equation*}
One could also determine the transformation rules of higher total $t$-derivatives of $q$ in a recursive way, if needed.

\subsection{Induced transformations and symmetries of kinematic functions}

According to the above transformation rules, while the triple $(t,q,\dot q)$ is transformed into $(t_\varepsilon,q_\varepsilon,\dot q_\varepsilon)$ by $\Phi$ along an evolution, the value of a smooth kinematic function $F_1$ of the time, position, and velocity, undergoes the transformation
\begin{equation*}
F_1(t,q,\dot q)\longrightarrow F_1(t_\varepsilon,q_\varepsilon,\dot q_\varepsilon)=F_1(t,q,\dot q)+\varepsilon\mathsf X^{[1]}(F_1)(t,q,\dot q),
\end{equation*}
where 
\begin{equation}
\mathsf X^{[1]}=\tau\,\frac{\partial}{\partial t}+\xi^i\frac{\partial}{\partial q^i}+(\dot\xi^i-\dot q^i\dot\tau)\frac{\partial}{\partial\dot q^i}\label{firstprolongation}
\end{equation}
is the so-called first prolongation \cite{Olver} of $\mathsf X $ especially built to act upon such functions. Here again, the transformation is a symmetry of $F_1$ if it leaves invariant the values of that function, i.e.\ if $\mathsf X^{[1]}(F_1)$ vanishes identically. One could also prolong $\mathsf X $ to deal with kinematic functions depending on $\ddot q$ and possibly higher total $t$-derivatives of $q$. However, it will not be necessary for our purpose. 

\subsection{Adapted coordinate systems}

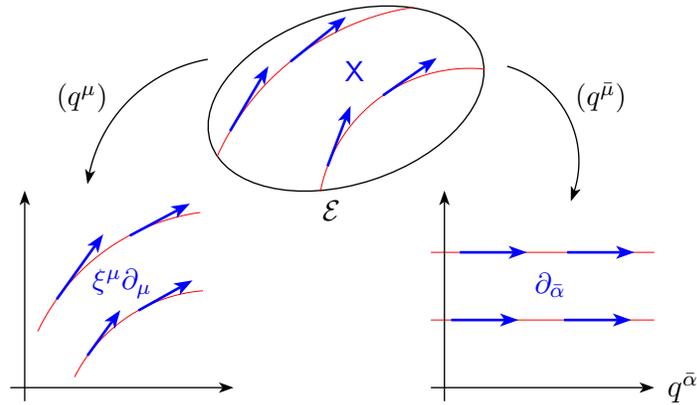
\begin{figure}
\centering
\psset{xunit=.6cm,yunit=.6cm,algebraic=true,dotstyle=o,dotsize=3pt 0,linewidth=0.5pt,arrowsize=3pt 2,arrowinset=0.25}
\begin{pspicture*}(0.05,0.)(15.3,8.95)
\parametricplot[linewidth=.3pt,linecolor=red]{1.7093114184455844}{2.696464557035488}{1*4.71*cos(t)+0*4.71*sin(t)+4.94|0*4.71*cos(t)+1*4.71*sin(t)+-0.38}
\parametricplot[linewidth=.3pt,linecolor=red]{1.651358282808984}{2.6743618432795206}{1*3.5*cos(t)+0*3.5*sin(t)+4.63|0*3.5*cos(t)+1*3.5*sin(t)+-0.94}
\psline{->}(0.08,0.4)(5.03,0.4)
\psline{->}(9.4,0.4)(14.35,0.4)
\uput[0](14.35,0.4){$q^{\bar\alpha}$}
\rput{18.19}(7.52,6.81){\psellipse(0,0)(3.17,1.91)}
\psline[linewidth=.3pt,linecolor=red](9.4,3.4)(14.35,3.4)
\psline[linewidth=.3pt,linecolor=red](9.4,1.9)(14.35,1.9)
\psline{->}(0.4,0.1)(0.4,4.77)
\psline{->}(9.7,0.1)(9.7,4.77)
\psline[linewidth=1pt,linecolor=blue]{->}(1.11,2.36)(2.14,3.8)
\psline[linewidth=1pt,linecolor=blue]{->}(2.73,3.77)(4.06,4.48)
\psline[linewidth=1pt,linecolor=blue]{->}(1.79,1.11)(2.52,2.12)
\psline[linewidth=1pt,linecolor=blue]{->}(2.92,2.11)(4.12,2.78)
\parametricplot[linewidth=.3pt,linecolor=red]{1.7235674377538832}{2.718741414308379}{1*5.72*cos(t)+0*5.72*sin(t)+9.88|0*5.72*cos(t)+1*5.72*sin(t)+3.18}
\parametricplot[linewidth=.3pt,linecolor=red]{1.472907948870405}{2.949113549971398}{1*3.37*cos(t)+0*3.37*sin(t)+10.26|0*3.37*cos(t)+1*3.37*sin(t)+4.11}
\psline[linewidth=1pt,linecolor=blue]{->}(4.95,6.09)(5.8,7.53)
\psline[linewidth=1pt,linecolor=blue]{->}(6.29,7.63)(7.52,8.62)
\psline[linewidth=1pt,linecolor=blue]{->}(7.11,5.31)(7.62,6.65)
\psline[linewidth=1pt,linecolor=blue]{->}(8.34,6.88)(9.53,7.71)
\psline[linewidth=1pt,linecolor=blue]{->}(9.85,1.9)(11.34,1.9)
\psline[linewidth=1pt,linecolor=blue]{->}(12.35,1.9)(13.84,1.9)
\psline[linewidth=1pt,linecolor=blue]{->}(10.04,3.4)(11.53,3.4)
\psline[linewidth=1pt,linecolor=blue]{->}(12.42,3.4)(13.91,3.4)
\parametricplot{1.7448508960176878}{2.9743639964300783}{1*3.26*cos(t)+0*3.26*sin(t)+5.02|0*3.26*cos(t)+1*3.26*sin(t)+4.47}
\psline{->}(1.81,5.01)(1.79,4.91)
\parametricplot{-0.36142931821872804}{1.293687305190943}{1*2.19*cos(t)+0*2.19*sin(t)+10.49|0*2.19*cos(t)+1*2.19*sin(t)+5.42}
\psline{->}(12.54,4.64)(12.5,4.53)
\begin{small}
\rput[bl](1.1,6.5){$(q^\mu)$}
\rput[bl](12.6,6.5){$(q^{\bar\mu})$}
\rput[bl](7.5,7.2){\color{blue}$\mathsf X$}
\rput[bl](7.,4.1){$\mathcal E$}
\rput[bl](1.9,2.4){\color{blue}$\xi^\mu\partial_\mu$}
\rput[bl](11.7,2.4){\color{blue}$\partial_{\bar\alpha}$}
\end{small}
\end{pspicture*}
\caption{Schematic illustration of the vector field $\mathsf X$ over $\mathcal E$ and its representatives with respect to an arbitrary coordinate system $(q^\mu)$ and an adapted one $(q^{\bar\mu})$. In the latter, the integral curves of $\mathsf X$ coincide with the coordinate lines of $q^{\bar\alpha}$.}\label{fig:coordinates}
\end{figure}

For obvious practical reasons, once we have chosen a coordinate system $(q^\mu)$, each individual coordinate $q^\mu$ is tacitly identified with the $\mu$-th coordinate function which maps events to their value of $q^\mu$. From the transformation \eref{transfo} regarding the coordinates (which formally reads $q^\mu\to q^\mu+\varepsilon\xi^\mu$), and from the general transformation rule \eref{transpoint} of the point functions, one readily deduces that applying $\mathsf X$ to the coordinate functions $q^\mu$ yields the equalities $\mathsf X(q^\mu)=\xi^\mu$. This observation allows us to write the components of $\mathsf X$ in an arbitrary system $(q^\mu)$ as $(\xi^\mu)=(\mathsf X(q^\mu))$. In another system $(q^{\mu'})$ they will be $(\xi^{\mu'})=(\mathsf X(q^{\mu'}))$, that is, $(\xi^{\mu'})=(\xi^\nu\partial_\nu q^{\mu'})$. This last equality encodes, as expected, the contravariant transformation rule of the components of vector fields.

According to the elementary theory of differential geometry \cite{Spivak}, it is always possible to define an extended coordinate system $(q^{\bar\mu})$ reducing the transformation $\Phi$ to a mere rigid translation of magnitude $\varepsilon$ along one of the coordinates, $q^{\bar\alpha}$ say (see figure \ref{fig:coordinates}). In such a system which is said to be \emph{adapted} to $\Phi$, the total derivatives of the coordinates $q^{\bar i}$ with respect to $\bar t=q^{\bar 0}$ remain invariant at all orders, thus $\mathsf X $ and its prolongations simply coincide with the partial derivative $\partial_{\bar\alpha}$. Hence, saying that $\Phi$ is a symmetry of a point or kinematic function $F$ amounts to saying that $F$ does not depend on the coordinate $q^{\bar\alpha}$ in the adapted system (although it may depend on its total derivatives with respect to $\bar t$ if $q^{\bar\alpha}\ne\bar t$). Alternatively stated, that kind of symmetry allows to reduce by one the number of variables necessary to describe $F$.

\section{Noether's theory and the Rund-Trautman function}\label{sec:Noether}

\subsection{Introduction of the Rund-Trautman function}

Suppose that the dynamics of the system derives from the Hamilton's principle applied to an action functional \cite{Goldstein}
\begin{equation*}
S(\mathcal C)=\int_{t_1}^{t_2} L[q(t)]\,\diff t,
\end{equation*} 
where $L$ is a a smooth Lagrangian of the first order and $[q(t)]$ a shorthand notation for the triple of arguments $(t,q(t),\dot q(t))$. It amounts to say that the motions are the evolutions satisfying the Euler-Lagrange equations
\begin{equation*}
\mathsf E_i(L)=0\qquad(i=1,\dots,n)
\end{equation*}
where
\begin{equation}
\mathsf E_i=\frac{\partial}{\partial q^i}-\frac{\diff}{\diff t}\frac{\partial}{\partial\dot q^i}\label{ELop}
\end{equation}
is the $i$-th Euler-Lagrange operator with respect to the used coordinate system. 

Under $\Phi$, the value of the action transforms as
\begin{eqnarray}
S(\mathcal C)\longrightarrow S(\mathcal C_\varepsilon)&=\int_{t_{1\varepsilon}}^{t_{2\varepsilon}}L[q_\varepsilon(t_\varepsilon)]\,\diff t_\varepsilon\,\label{transformedS}
\end{eqnarray}
where $[q_\varepsilon(t_\varepsilon)]$ stands for the triple of arguments $(t_\varepsilon,q_\varepsilon(t_\varepsilon),\dot q_\varepsilon(t_\varepsilon))$. To the first order in $\varepsilon$, the induced variation of the action is (see \ref{app:var})
\begin{equation}
\delta S(\mathcal C)=S(\mathcal C_\varepsilon)-S(\mathcal C)=\varepsilon\int_{t_1}^{t_2}\bigg(\mathsf X^{[1]}(L)+\dot\tau L\bigg)[q(t)]\,\diff t.\label{transS}
\end{equation}
Now, let us introduce some smooth point function $B$ (given up to a meaningless additive constant) and contemplate the difference
\begin{equation}
\mathcal D(\mathcal C)=\Big[\varepsilon B(t,q(t))\Big]_{t_1}^{t_2}-\delta S(\mathcal C)=\varepsilon\int_{t_1}^{t_2}R[q(t)]\,\diff t,\label{difference}
\end{equation}
where has been introduced the kinematic function
\begin{equation}
R(t,q,\dot q)=\dot B-\mathsf X^{[1]}(L)-\dot\tau L\label{RT}
\end{equation}
that will be called the \emph{Rund-Trautman function} defined by the Lagrangian $L$, the transformation $\Phi$ and the \emph{boundary term} $B$. Rearranging its right-hand side (see \ref{app:I}), the last equation can be rewritten 
\begin{equation}
R=\dot I-(\xi^i-\dot q^i\tau)\mathsf E_i(L),\label{relation}
\end{equation} 
where was introduced the quantity
\begin{equation}
I(t,q,\dot q)= B+H\tau-p_i\xi^i=B-p_\mu\xi^\mu,\label{FI}
\end{equation}
with $p_i=\partial L/\partial{\dot q^i}$ and $p_0=-H=L-p_i\dot q^i$ the components of the extended momentum. From \eref{relation}, one sees that $R$ is the rate of change of $I$ along the motions.

The fact that the left-hand side of \eref{difference} has a coordinate-free meaning suffices to say that $R\diff t$ in the right-hand side is invariant under a change of extended coordinates. It is also the case of $I$ since $B$ and the contraction $p_\mu\xi^\mu$ are scalars.

\subsection{Harmonization with the Lagrangian gauge freedom}\label{subsec:gauge}

One knows that the dynamics is invariant under the addition of a total differential $\diff G$ to the form $L\diff t$, where $G$ is a point function. This addition corresponds to the gauge transformation $L\to \widetilde L= L+\dot G$ of the Lagrangian. Accordingly, the action gauge-transforms as
\begin{equation*}
S(\mathcal C)\longrightarrow \widetilde S(\mathcal C)=\int_{t_1}^{t_2}\widetilde L[q(t)]\,\diff t=S(\mathcal C)+\Big[G(t,q(t))\Big]_{t_{1}}^{t_{2}}
\end{equation*}
and its variation under $\Phi$ as
\begin{equation*}
\delta  S(\mathcal C)\longrightarrow\delta \widetilde S(\mathcal C)=\widetilde S(\mathcal C_\varepsilon)-\widetilde S(\mathcal C).
\end{equation*}
As is shown in \ref{app:gauge}, one obtains, to the first order in $\varepsilon$:
\begin{equation}
\delta  \widetilde S(\mathcal C)=\delta  S(\mathcal C)+\varepsilon\,\Big[\mathsf X (G)(t,q(t))\Big]_{t_1}^{t_2}\,.\label{vardeltaS}
\end{equation}
The difference $\mathcal D(\mathcal C)$ is rendered gauge-invariant if one endows with $B$ the compensating gauge transformation law
\begin{equation*}
B\longrightarrow\widetilde B=B+\mathsf X(G).
\end{equation*}
Indeed, it is easily verified that
\begin{equation*}
\Big[\varepsilon \widetilde B(t,q(t))\Big]_{t_1}^{t_2}-\delta\widetilde S(\mathcal C)=\Big[\varepsilon B(t,q(t))\Big]_{t_1}^{t_2}-\delta S(\mathcal C).
\end{equation*}
Consequently, the Rund-Trautman function is gauge invariant ($\widetilde R=R$), as well as $I$, according to the gauge transformation rule of the extended momenta:
\begin{equation*}
\widetilde I=\widetilde B-\tilde p_\mu\xi^\mu=\widetilde B-(p_\mu+\partial_\mu G)\xi^\mu=\widetilde B-p_\mu\xi^\mu-\mathsf X(G)=B-p_\mu\xi^\mu=I.
\end{equation*}

\subsection{Trivialization of the formalism through adapted gauge and coordinates}\label{subsec:triv}

Locally, one can always choose a point function $G$ verifying the gauge condition $B+\mathsf X (G)=0$ in order to cancel the boundary term. Such a gauge will be said to be adapted to the couple formed by the transformation and the boundary term. It can be most easily done by using an adapted coordinate system $(q^{\bar\mu})=(\bar t,q^{\bar i})$ such that $\mathsf X=\partial_{\bar\alpha}$. Indeed, the gauge condition becomes $B+\partial_{\bar\alpha}G=0$ and an integral of this expression with respect to $q^{\bar\alpha}$ suffices to obtain a suitable point function $G$.

Once this is done, the Lagrangian $\overline L$ expressed in the adapted system and defined by $\overline L\diff\bar t=L\diff t+\diff G$ is given in an adapted gauge as well. The Rund-Trautman function is now simply $\overline R=-\partial_{\bar\alpha}\overline L$ and $I$ reduces to
\begin{equation*}
I=-\overline p_{\bar\alpha}=-\frac{\partial \overline L}{\partial\mathring q^{\bar\alpha}}\,,
\end{equation*}
where, in order to avoid any confusion, the empty bullet symbolizes the derivation with respect to the adapted time $\bar t$. Hence, equation \eref{relation} becomes, along the motions, tantamount to the Euler-Lagrange equation $\mathsf E_{\bar\alpha}(\overline L)=0$. Actually, $\overline R$ measures the dependence of the adapted Lagrangian $\overline L$ on $q^{\bar\alpha}$ as well as the rate of change of the momentum $\overline p_{\bar\alpha}$ along the motions (up to a sign).

\subsection{Symmetry or not symmetry}

One says that the transformation $\Phi$ is a Noether point symmetry (NPS) of the problem if there exists a boundary term $B$ such that $\mathcal D(\mathcal C)$ vanishes for any evolution $\mathcal C$. In this case, $R$ identically vanishes and thus the quantity $I$ is conserved along the motions. According to Paragraph \ref{subsec:gauge}, an NPS is as expected a gauge-invariant property generating a gauge-invariant constant of motion. In particular, an NPS leaves the action integral invariant in an adapted gauge and the symmetry is said to be strict in this case. 

If $\Phi$ is an NPS then its meaning becomes transparent as seen through the lens of the adapted Lagrangian $\overline L$ introduced in Paragraph \ref{subsec:triv}. Indeed, it says that $q^{\bar\alpha}$ is a cyclic coordinate\footnote{In addition, $p_{\bar\alpha}$ being a partial derivative of $\overline L$, it does not depend on $q^{\bar\alpha}$ either. This point demonstrates that an NPS $\Phi$ is a symmetry of the constant of motion $I$.}, or equivalently that its conjugate momentum is conserved.

In general, however, $R$ is interpretable as the rates of change of both the `asymmetricity' and $I$. Therefore, seeking `almost' vanishing Rund-Trautman functions can be a good approach to find `almost' conserved quantities. In the adiabatic context, the almostness in question will become clearer in Section \ref{sec:applications}. But, before, we apply the theory discussed in this section to the general one-dimensional Lagrangian problems.

\section{Application to one-dimensional problems}\label{sec:applications}

\subsection{General aspects}

We consider in this section the standard Lagrangian of a unit-mass particle experiencing a potential $V(t,q)$ along a straight line:
\begin{equation*}
L=\frac12\,\dot q^2-V(t,q).
\end{equation*}
Let us introduce the generator $\mathsf X=\tau\partial_t+\xi\partial_q$ of a point transformation $\Phi$ as well as a boundary term $B$. The corresponding Rund-Trautman function \eref{RT} takes the form
\begin{equation*}
R=R_3(t,q)\dot q^3+R_2(t,q)\dot q^2+R_1(t,q)\dot q+R_0(t,q),
\end{equation*} 
where
\begin{equation}
\begin{array}{ll}
R_3=\displaystyle\frac12\,\partial_q\tau,\phantom{\bigg|} & R_2=\displaystyle\frac12\,\partial_t\tau-\partial_q\xi, \\
R_1=\partial_qB+V\partial_q\tau-\partial_t\xi,\phantom{\bigg|}\qquad &
R_0=\partial_t(B+\tau V)+\xi\partial_qV.
\end{array}\label{Rs}
\end{equation}
The three quantities $R_3$, $R_2$ and $R_1$, taken in this order, are found to identically vanish iff $\tau$, $\xi$ and $B$ have the form
\begin{equation}
\begin{array}{ll}
\tau=\tau(t), & \xi=\displaystyle\frac12\,\dot\tau(t) q+\psi(t), \\
B=\displaystyle\frac14\,\ddot\tau(t) q^2+\dot\psi(t) q+\chi(t),
\end{array}\label{taupsichi}
\end{equation}
whatever $V$ may be. From now on, $\tau$, $\xi$ and $B$ have these forms in which the three functions $\tau(t)$, $\psi(t)$ and $\chi(t)$ are arbitrary but fixed. Therefore, the Rund-Trautman function $R$ reduces to the point function $R_0$. 

We will restrict ourselves to the case $\tau\ne 0$, i.e.\ to asynchronous transformations. Reversing $\Phi$ if necessary, one can suppose the transformation future-oriented ($\tau>0$) without loss of generality. It is shown in \ref{app:changevar} that the change of extended coordinates $(t,q)\to(T,Q)$, with
\begin{equation}
T=\int^t\frac{\diff t}{\tau}\qquad\text{and}\qquad Q=\frac{q}{\sqrt{\tau}}-\int^t\frac{\psi}{\tau^{3/2}}\,\diff t,\label{TQ}
\end{equation}
reduces $\Phi$ to the translation $(T,Q)\to(T+\varepsilon,Q)$, and its generator $\mathsf X$ to $\partial_T$. Using the adapted coordinates, one obtains after some straightforward but lengthy calculations (see \ref{app:V}), that the expression of $R_0$ in \eref{tauR0} is equivalent to the existence of a function $W(Q)$ such that
\begin{equation}
V=\frac{1}{\rho^2}\,W(Q)-\frac{\ddot\rho}{2\rho}\, q^2-\frac{1}{\rho}\frac{\diff(\rho^2\dot\alpha)}{\diff t}\,q+\beta+\frac{1}{\rho^2}\int^T\hspace{-1mm}\rho^2R_0\,\diff T',\label{Valternatif}
\end{equation} 
where
\begin{equation}
\rho=\sqrt{\tau}\,,\qquad\alpha=\int^t\frac{\psi}{\tau^{3/2}}\,\diff t\qquad\text{and}\qquad \beta=\frac{\psi^2}{2\tau^2}-\frac{\chi}{\tau}\label{rho_beta}
\end{equation}
are functions of the time only. The function $W(Q)$ is defined up to a meaningless additive constant since an alteration $W(Q)\to W(Q)+\text{cst.}$ can be compensated by a redefinition of $\chi$ through $\chi\to \chi+\text{cst}$. Actually, $\chi$ has no physical meaning, its role is only to ensure the gauge symmetry according to which adding to the potential an explicit function of the time only does not affect the dynamics.

Now, it is shown in \ref{app:Lambda} that the gauge condition $B+\mathsf X(G)=0$ is fulfilled if one chooses
\begin{equation}
G=-\frac{\dot\rho}{2\rho}\,q^2-\rho\dot\alpha q+\int^t\Bigg(\frac12\,\rho^2\dot\alpha^2+\beta\Bigg)\diff t.\label{Lambda}
\end{equation}
Then, following the method described in Paragraph \ref{subsec:triv} (see \ref{app:L}), one obtains the adapted Lagrangian
\begin{equation}
\overline L(T,Q,\mathring Q)=\frac 12\, \mathring Q^2-W(Q)-\int^T\hspace{-1mm}\rho^2 R_0\,\diff T,\label{newlag}
\end{equation}
where the empty bullet symbolizes the total derivation with respect to the new time $T$. The integrand of the indefinite integral in \eref{newlag} is actually the new Rund-Trautman function and is, as expected, the opposite of the partial derivative of the new Lagrangian with respect to $T$. Moreover, the new energy function coincides with $I$: 
\begin{equation*}
I=\overline H=\frac 12\, \mathring Q^2+W(Q)+\int^T\hspace{-1mm}\rho^2 R_0\,\diff T.
\end{equation*}

From the above discussion, we deduce that the variational problem admits asynchronous NPS (ANPS) iff the potential can be written in a form
\begin{equation}
V=\frac{1}{\rho^2}\,W\Bigg(\frac{q}{\rho}-\alpha\Bigg)-\frac{\ddot\rho}{2\rho}\, q^2-\frac{1}{\rho}\frac{\diff(\rho^2\dot\alpha)}{\diff t}\,q+\beta,\label{sym_V}
\end{equation}
with $\rho>0$, $\alpha$ and $\beta$ three functions of the time. In this case, by inverting the equalities \eref{rho_beta} one constructs the functions $\tau$, $\psi$, $\chi$ which define, through \eref{taupsichi}, the vector field of the ANPS along with the boundary term. In the adapted coordinates \eref{TQ}, the problem is transformed into the conservative problem of a particle experiencing a time-independent potential $W$. In this new viewpoint, all becomes transparent: the symmetry ($T$-invariance) and the first integral (the new energy).

\subsection{Noether point symmetries of quadratic potentials}\label{subsecquadra}

Let us apply the above considerations to the frequently encountered quadratic potentials
\begin{equation}
V=\frac12\,a(t)q^2+b(t)q,\label{quadratic}
\end{equation}
where $a(t)$ and $b(t)$ are some functions of the time. Since the adapted coordinate $Q$ is necessarily linear in $q$, the only candidates to the function $W(Q)$ in \eref{sym_V} have clearly the form
\begin{equation*}
W(Q)=\frac 12\,C_2Q^2+C_1Q+C_0\,,
\end{equation*}
where $C_2$, $C_1$ and $C_0$ are constants. Now, substituting $W(Q)$ in \eref{sym_V} by the above expression and identifying the result with \eref{quadratic}, one obtains that $\rho>0$, $\alpha$ and $\beta$ must verify
\numparts\begin{eqnarray}\label{system}
\rho^3(\ddot\rho+a\rho)&=&C_2\,,\label{systemA}\\
\rho^3\bigg[\frac{\diff^2}{\diff t^2}\big(\alpha\rho\big)+a\alpha\rho+b\bigg]&=&C_1\,,\\
\beta\rho^2+\frac12\,C_2\alpha^2-C_1\alpha &=&C_0\,.
\end{eqnarray}\endnumparts

Fixing the three constants in the right-hand sides at arbitrary values, any solution of the differential equations \cref{system} gives an ANPS with its boundary term. Actually, the system of equations \cref{system} says that the left-hand sides must be constant, i.e.\ that their derivatives are zero. Differentiating them, one obtains
\numparts\begin{eqnarray}\label{eqtaupsichi}
\frac14\,\dddot\tau+a\dot\tau+\frac12\,\dot a\tau&=&0,\label{eqtau}\\
\ddot\psi+a\psi+\frac32\,b\dot\tau+\dot b\tau&=&0,\label{eqpsi}\\
\dot\chi+b\psi&=&0.\label{eqchi}
\end{eqnarray}\endnumparts
But these equations are precisely the necessary and sufficient conditions for $\Phi$ to be an NPS of $S$ with boundary term $B$. Indeed, inserting the expression \eref{quadratic} of the potential in the expression of $R_0$ in \eref{Rs} and taking into account the relations \eref{taupsichi}, one has
\begin{equation*}
R_0=\bigg[\frac14\,\dddot\tau+a\dot\tau+\frac12\,\dot a\tau\bigg]q^2+\bigg[\ddot\psi+a\psi+\frac32\,b\dot\tau+\dot b\tau\bigg]q+\big[\dot\chi+b\psi\big].
\end{equation*}

The general solution of \eref{eqtau}, seen as a differential equation in $\tau$, depends linearly on three parameters, following which the general solution of \eref{eqpsi}, seen as a differential equation in $\psi$, depends linearly on two supplementary parameters. Alternatively stated, the Noether point symmetry group of $S$ for quadratic potentials is five dimensional \cite{Prince} and generated by three asynchronous transformations $(\tau\ne 0)$ and two synchronous others $(\tau=0)$. It has been shown in Reference \cite{Leone2017} that the last two ones actually manifest the linearity of the equation of motion.

Conversely, it is also clear that the quadratic potentials are the only ones which allow for synchronous NPS (SNPS)\footnote{If $\tau=0$ then, taking into account \eref{taupsichi}, the expression of $R_0$ in \eref{Rs} becomes $-\psi\partial_qV-\ddot\psi q-\dot\chi$ and it is clear that it can identically vanish only if $V$ is a quadratic potential.}. While ANPS lead to first integrals quadratic in the velocities which are `energy-like', SNPS lead to first integrals linear in the velocities which are `momentum-like'.

\subsection{The particular case of the time-dependent harmonic oscillator}\label{HO}

The harmonic oscillator with time-dependent frequency $\omega(t)$ deserves a special attention. Here, the potential $V$ has the form \eref{quadratic} in which $a(t)=\omega^2(t)$ and $b(t)=0$. For $\Phi$ to be an ANPS, it suffices to take $\psi=\chi=0$, and $\tau$ a solution of \eref{eqtau} or, equivalently, $\rho$ a solution of Ermakov's equation \cite{Ermakov} \eref{systemA} for some value of $C_2$. The conserved quantity $I$ is now the so-called Ermakov-Lewis invariant \cite{Ermakov,Lewis1967} 
\begin{equation}
I=\overline H=\frac12\,\mathring Q^2+\frac12\,C_2Q^2=\frac12(\rho\dot q-\dot\rho q)^2+\frac{C_2}{2\rho^2}\,q^2.\label{ELinv}
\end{equation} 
The new problem is then easily solved in $Q(T)$, and $q(t)$ follows immediately. In particular, if $C_2>0$, the general solution reads
\begin{equation}
q(t)=\sqrt{2I\tau}\,\cos\Bigg(\sqrt{C_2}\int_0^t\frac{\diff t}{\tau}+\varphi_0\Bigg).\label{general_solution}
\end{equation}

\subsubsection*{An example.}

An interesting case occurs when the period of the oscillator is a quadratic function of the time, that is, when the frequency has the form
$\omega(t)=(1+2\gamma t+\delta t^2)^{-1}$ in a suitable unit of time, $\gamma$ and $\delta$ being two constants. Indeed, since Equation \eref{eqtau} can be rewritten
\begin{equation}
\frac14\,\dddot\tau+\omega\,\frac{\diff}{\diff t}\big(\omega\tau\big)=0,\label{eqtauHO}
\end{equation}
it is clear that setting $\tau=\omega^{-1}$ allows to cancel separately both the terms of the left-hand side. Hence, $\tau=\omega^{-1}$ is a solution of \eref{eqtauHO} coming with the integrating constant $C_2=1-\gamma^2+\delta$. Historically, that profile of frequency was for example considered by Fock in a quantum mechanical context \cite{Fock}.

\section{Adiabaticity and almost Noether point symmetries}\label{sec:adiabatic}

\subsection{General considerations}

As we saw above, the Rund-Trautman function \eref{RT} measures the rate of change of the quantity $I$ in \eref{FI} along the motions. If $\Phi$ is not an NPS of $S$ with boundary term $B$, it can nevertheless be expected that it is `almost' such a symmetry, the almostness in question being quantifiable with respect to some quantities that we have every reason to regard them as small. In the usual perturbation theory, they measure the weakness of the couplings between a system and its environment. In the adiabatic theory, they are the slow rate of change of time-dependent parameters on which the dynamics depends.

For simplicity, suppose that the system depends on a single tunable parameter $\lambda$ and that the observer wants to make it pass from an initial value $\lambda_0$ to a final value $\lambda_1$. To this end, he chooses which evolution pattern $\lambda(s)$ he will make the parameter follow to reach $\lambda_1=\lambda(1)$ from $\lambda_0=\lambda(0)$. At this stage, $s\in[0,1]$ is only an abstract evolution parameter whose variations $\Delta s$ will be proportional to $\Delta t$. Then, he still needs to decide how long the process will take, or, putting it differently, to fix the proportionality factor $\eta=\Delta s/\Delta t>0$ at a certain value. Hence, if the beginning of the evolution is taken at $t=0$ then one has $s=\eta t$, the final time is $t=\eta^{-1}$, and at any intermediate instant $t$ the value of the parameter is $\lambda(\eta t)$. The more $\eta$ is small, the more $\lambda$ evolves slowly, and the adiabatic regime is reached in the limit $\eta\to0$. Obviously, it is only an unattainable horizon and one speaks of adiabaticity as soon as $\eta$ can be considered as very small as compared to the typical frequencies of the dynamics \cite{Goldstein,Arnold}. For simplicity again, we will remain in the situation where the dynamics is entirely embodied in a Lagrangian whose explicit time dependence is, by assumption, only realized via $\lambda$ (and not on its derivatives).

Hereafter, a function $F(s,q,\dot q;\eta)$ will be said to be formally of the order $\nu$ ($\nu\geqslant 0$) if, when $\eta$ approaches 0, the ratio $F/\eta^\nu$ converges to a finite function of $s$, $q$, $\dot q$. The reason why $s$ is privileged over $t$ is clear since $s$ is bounded unlike $t$ which goes to infinity when $\eta$ approaches 0. For the same reason, along the motion, the $t$-derivative of $q$ is expected to remain bounded whereas its $s$-derivative would reach infinite values in the same limit $\eta\to 0$. 

Now, if a couple $(\Phi,B)$ is such that the Rund-Trautman function \eref{RT} is formally of some order $\nu+1$, then, the rate of change of $I$ along the motion is formally of the order $\nu+1$ with respect to $t$, and $\nu$ with respect to $s$. Hence, it is expected that the discrepancy between the initial and final values of $I$ along a motion taking place between $t=0$ and $t=\eta^{-1}$ is an $\mathrm O(\eta^{\nu})$, i.e.\ that $I$ is an adiabatic invariant of the $\nu$-th order. For the sake of illustration, we will develop this idea in the paradigmatic problem of an harmonic oscillator with a slowly varying frequency.

\subsection{The time-dependent harmonic oscillator}

We consider an harmonic oscillator with a slowly varying frequency $\omega(\eta t)$, where $\eta$ is very small as compared to the values taken by $\omega$ between $t=0$ and $t=\eta^{-1}$. Treating this example seems at first glance surprising since it is nothing but an application of the Paragraph \ref{HO} from which one formally knows constants of motion which are in some sense adiabatic invariants to all orders. But, in general, the form of $\omega(\eta t)$ does not allow for a solution of \eref{eqtauHO} having a simple analytic expression. This is the reason why it is more useful to seek approximate solutions. An integrating constant $C_2=1$ would transform the problem into the one of an harmonic oscillator with unit frequency. So, let us search for an approximate solution $\rho=\sqrt\tau$ of
\begin{equation*}
\rho^3(\ddot\rho+\omega^2(\eta t)\rho)=1.
\end{equation*}
Denoting the total derivative with respect to $s=\eta t$ by a prime, the left-hand side can be transformed to yield
\begin{equation}
\frac{\eta^2}{2}\Bigg(\tau\tau''-\frac12\,\tau'^2\Bigg)+\omega^2(s)\tau^2=1.\label{etatau}
\end{equation}
Then, if we insert a perturbative expansion
\begin{equation*}
\tau(s)=\tau_0(s)+\eta^2\tau_1(s)+\dots+\eta^{2k}\tau_{k}(s)+\dots
\end{equation*}
in \eref{etatau}, one obtains after an identification of its two sides the formal expressions
\begin{eqnarray}
\tau_0&=\omega^{-1}\nonumber\\
\tau_1&=\frac{1}{4\omega}\Bigg(\frac12\,{\tau'_0}^2-\tau_0\tau''_0\Bigg)\label{composantes}\\
\tau_{k}&=\frac{1}{4\omega}\sum_{i=1}^k\Bigg(\frac12\,\tau'_{i-1}\tau'_{k-i}-\tau_{i-1}\tau''_{k-i}\Bigg)-\frac{\omega}{2}\sum_{i=1}^{k-1}\tau_{i}\tau_{k-i}\qquad(k\geqslant 2)\nonumber
\end{eqnarray}
The $\tau_k$s are thereby deduced step by step provided that $\omega$ is sufficiently differentiable (each $\tau_k$ is expressible in terms of $\omega^{-1}$ and its $k$ first derivatives). The quantity $I$ thus admits a formal expansion
\begin{equation}
I=I_0+\eta I_1+\eta^2I_2+\dots+\eta^kI_k+\dots\label{I_expansion}
\end{equation}
in which the components of even orders are
\begin{equation*}
I_0=\frac{H}{\omega}\qquad,\qquad I_{2k}=H\tau_k+\frac14\,q^2\tau''_{k-1}\qquad (k\geqslant 1)
\end{equation*}
while the components of odd orders are
\begin{equation*}
I_{2k+1}=-\frac12\,q\dot q\tau'_k\qquad(k\geqslant 0)
\end{equation*}
A truncation
\begin{equation*}
I_{(k)}=I_0+\eta I_1+\eta^2I_2+\dots+\eta^kI_k\qquad(k\geqslant 0)
\end{equation*} 
of $I$ is formally of the order $k$ since its $s$-derivative, depending on its parity, is given by
\begin{equation*}
I'_{(2k)}=\frac12\,\eta^{2k}\big(\dot q^2-\omega^2q^2\big)\tau'_k\qquad,\qquad I'_{(2k+1)}=-\frac12\,\eta^{2k+1}q\dot q\tau''_k
\end{equation*}
along the motion. Consider an increment $\Delta s$ such that $\eta\omega\ll\Delta s\ll 1$ on $[s,s+\Delta s]$. By the assumption $\Delta s\ll 1$, the slow quantities $\tau'_k$ and $\tau''_k$ are supposed to be almost constant on this interval. The derivatives of the truncations $I_{(k)}$ along the motion are thus, at $s$, approximately
\begin{equation*}
I'_{(2k)}\approx\frac12\,\eta^{2k}\big\langle\dot q^2-\omega^2q^2\big\rangle\tau'_k\qquad,\qquad I'_{(2k+1)}\approx-\frac12\,\eta^{2k+1}\big\langle q\dot q\big\rangle\tau''_k,
\end{equation*}  
where the averages are taken over $[s,s+\Delta s]$. Moreover, according to the general solution \eref{general_solution}, since $\Delta t=\eta^{-1}\Delta s\gg\omega^{-1}$, the phases of $\omega q$ and $\dot q$ highly oscillate between $s$ and $s+\Delta s$ whereas their amplitudes are, to the lowest order, $\sqrt{2H}+\mathrm O(\eta)$. Consequently, the averages $\langle \dot q^2- \omega^2q^2\rangle$ and $\langle q\dot q\rangle$ are approximately of the order $\mathrm O(\eta)$ and the truncations $I_{(k)}$ are adiabatic invariants of an order close to $k+1$. 

\subsection{A first explicit example: the frequency following a power law}

Suppose that the frequency profile have the form $\omega(s)=(1+\gamma s)^{r-1}$, with $\gamma$ and $r$ two constants. Using the recursive scheme \eref{composantes}, it can be easily verified that the $\tau_k$s have a certain form
\begin{equation*}
\tau_k=a_k\gamma^{2k}(1+\gamma s)^{1-(2k+1)r}
\end{equation*} 
in which the $a_k$s are coefficients ($a_0=1$). We can express $a_k$ in terms of $a_{k-1}$ and $k$ only by exploiting the equality \eref{eqtauHO} which brings to us the relations
\begin{equation}
\tau'''_{k-1}+4\omega(\omega\tau_k)'=0.\label{recursion_derivee}
\end{equation}
We find
\begin{equation*}
a_k=\frac{2k-1}{8k}\Big[1-(2k+1)^2r^2\Big]a_{k-1}\,.
\end{equation*}
The $a_k$s form a divergent hypergeometric sequence admitting the closed form
\begin{equation*}
a_k=\frac{(2k-1)!!}{k!\,8^k}\prod_{i=1}^k\Big[1-(2k+1)^2r^2\Big].
\end{equation*}
Finally, the components of the adiabatic expansion \eref{I_expansion} are
\begin{eqnarray*}
I_{2k}&=\frac{\tau_k}{2}\bigg[\dot q^2-\frac{(2k+1)r-1}{(2k-1)r+1}\,\omega^2q^2\bigg],\\
I_{2k+1}&=\frac{\tau_k}{2}\frac{(2k+1)r-1}{1+\gamma s}\,\gamma q\dot q.
\end{eqnarray*}

\subsection{A second explicit example: the frequency following an exponential law}

\begin{figure}
\centering
\includegraphics[scale=.75]{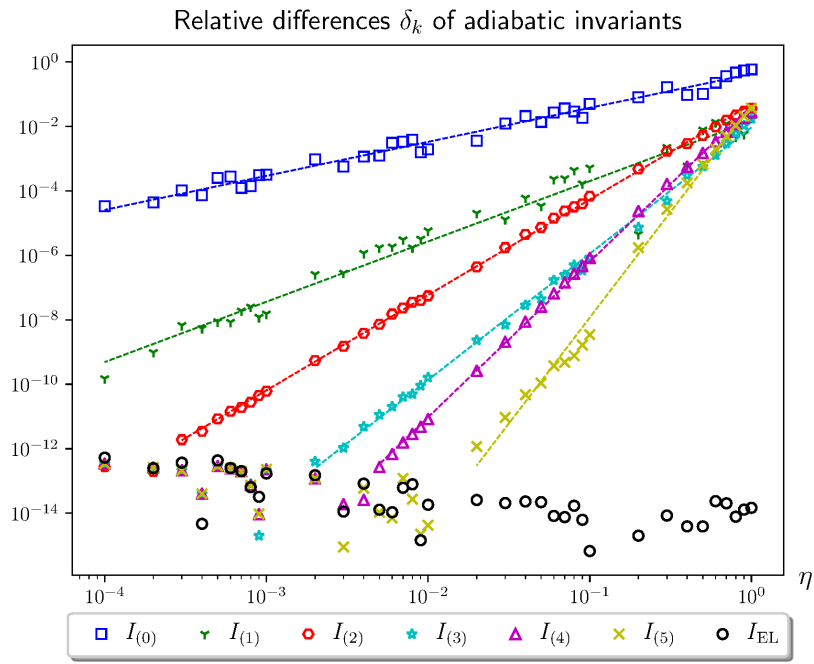}
\caption{Numerical analysis (color online). Relative changes of the six first adiabatic invariants --- and of an Ermakov-Lewis invariant $I_{\text{EL}}$ --- between the initial and final states, as functions of the parameter $\eta$, for the harmonic oscillator with a slowly varying frequency $\omega(\eta t)=2^{\eta t}$. The conjugate pairs $(q,\dot q)$ and $(\rho,\dot\rho)$ are integrated using the fourth-order symplectic Runge-Kutta-Nystr\"om algorithm developed by Calvo and Sanz-Serna \cite{Calvo}, with $q(0)=\rho(0)=1$ and $\dot q(0)=\dot\rho(0)=0$ as initial conditions ($t$-time step: $0.001$). Before becoming numerically noisy when too small, $\delta_k$ is approximately proportional to a power $\eta^{\nu_k}$ with $\nu_0=1.06\pm 0.03$, $\nu_1=1.87\pm0.07$, $\nu_2=2.96\pm0.01$, $\nu_3=3.92\pm0.04$, $\nu_4=4.83\pm0.03$ and $\nu_5=6.57\pm0.15$.}\label{courbes}
\end{figure}

We now consider a frequency profile $\omega(s)=\mathrm{e}^{\gamma s}$, where $\gamma$ is a constant. The recursion scheme \eref{composantes} reveals that the $\tau_k$s have a certain form
\begin{equation*}
\tau_k=\frac{b_k\gamma^{2k}}{\omega^{2k+1}}
\end{equation*}
in which the $b_k$s are coefficients ($b_0=1$). Applying the relations \eref{recursion_derivee}, the $b_k$s are again the elements of a diverging hypergeometric sequence here given by
\begin{equation*}
b_k=\frac{(-1)^k}{8^k}\frac{\big[(2k-1)!!\big]^3}{k!}\,.
\end{equation*}
The components of the adiabatic expansion \eref{I_expansion} are thus
\begin{eqnarray*}
I_{2k}&=\frac{\tau_k}{2}\bigg[\dot q^2-\frac{2k+1}{2k-1}\,\omega^2q^2\bigg],\\
I_{2k+1}&=\frac{\tau_k}{2}(2k+1)\gamma q\dot q.
\end{eqnarray*}

Let us end this example by a numerical experiment in the case of a doubling of the frequency realized exponentially, i.e.\ for $\omega(s)=2^s=\mathrm{e}^{\gamma s}$ with $\gamma=\log(2)$. For a set of values of $\eta$, the equation of motion is integrated with $q(0)=1$ and $\dot q(0)=0$ as initial conditions. Then, the final values of the six first adiabatic invariants $I_{(0)},\dots,I_{(5)}$ are compared with their initial values through the relative differences
\begin{equation*}
\delta_k=\left|\frac{ I_{(k)}(s=1)-I_{(k)}(s=0)}{I_{(k)}(s=0)}\right|.
\end{equation*}
We find that $\delta_{k}$ follows in good approximation a power law $\delta_k\propto\eta^{\nu_k}$ with $\nu_k$ close enough to $k+1$ (see figure \ref{courbes}). This illustrates the fact that $I_{(k)}$ is an adiabatic invariant of an order close to $k+1$.

\section{Final remarks}

In our approach of the adiabatic invariance from Noether's theory, we did not apply an averaging procedure on the Rund-Trautman identity, over a well-chosen slow variable \cite{Boccaletti}. Also, unlike the works of Neuenschwander \textit{et al} on the subject, we did not assume a certain type of potential admitting an exact Noether symmetry and a convenient Rund-Trautman function to work on \cite{Neuenschwander,NeuenschwanderAJP}. Let us add that the frequency of the harmonic oscillator was supposed differentiable a certain number of times. However, in general, a time-dependent frequency is used as a transition between two regimes and discontinuities in the derivatives must be taken into account at the junctions. The great interest of the historic adiabatic invariant $I_0$ lies in its insensitivity to them.

\appendix

\section{Some proofs of Section \ref{sec:events}}

\subsection{Transformation of an evolution for $|\varepsilon|$ small enough}\label{app:graph}

The function $u\colon t\mapsto\tau(t,q(t))$ is smooth by assumption. Its derivative $\dot u$ is thus bounded on $[t_1,t_2]$. Let $\alpha$ be an upper bound of $|\dot u|$ on this interval. For an arbitrary value $\varepsilon$ of the parameter such that $|\varepsilon|<\alpha^{-1}$, one has
\begin{equation*}
\frac{\diff t_{\varepsilon}(t,q(t)))}{\diff t}=\frac{\diff}{\diff t}\bigg[t+\varepsilon\tau(t,q(t))\bigg]=1+\varepsilon\dot u(t)>1-|\varepsilon \dot u(t)|>0
\end{equation*}
for any time $t\in[t_1,t_2]$. Since $t_\varepsilon$ strictly increases with $t$ on this interval, the transformed curve of $\mathcal C$ is also the graph of an evolution.

\subsection{The transformation of the velocities under $\Phi$}\label{app:vel}

Inserting \eref{transgraph} in \eref{transvel}, one has, to the first order in $\varepsilon$:
\begin{eqnarray*}
\dot q^i_\varepsilon(t_\varepsilon)&=\frac{\diff}{\diff t}\bigg[q^i(t)+\varepsilon\,\xi^i(t,q(t))\bigg]\Bigg\slash\frac{\diff}{\diff t}\bigg[t+\varepsilon\,\tau(t,q(t))\bigg]\\
&=\Bigg[\dot q^i(t)+\varepsilon\,\frac{\diff\xi^i(t,q(t))}{\diff t}\Bigg]\Bigg\slash\Bigg[1+\varepsilon\,\frac{\diff\tau(t,q(t))}{\diff t}\Bigg]\\
&=\Bigg[\dot q^i(t)+\varepsilon\,\frac{\diff\xi^i(t,q(t))}{\diff t}\Bigg]\cdot \Bigg[1-\varepsilon\,\frac{\diff\tau(t,q(t))}{\diff t}\Bigg]\\
&=\dot q^i(t)+\varepsilon\,\Bigg[\frac{\diff\xi^i(t,q(t))}{\diff t}-\dot q^i(t)\,\frac{\diff\tau(t,q(t))}{\diff t}\Bigg].
\end{eqnarray*}
This is Formula \eref{velocities}.

\section{Some proofs of Section \ref{sec:Noether}}

\subsection{The variation ${\delta S(\mathcal C)}$ of the action under $\Phi$}\label{app:var}
Seeing $t_\varepsilon$, $q_\varepsilon(t_\varepsilon)$ and $\dot q_\varepsilon(t_\varepsilon)$ as functions of $t$ through the equalities \eref{transgraph} and \eref{velocities}, one first makes the change of variable of integration $[t_{1\varepsilon},t_{2\varepsilon}]\ni t_\varepsilon\to t\in[t_1,t_2]$ in \eref{transformedS}:
\begin{equation*}
S(\mathcal C_\varepsilon)=\int_{t_{1\varepsilon}}^{t_{2\varepsilon}}L[q_\varepsilon(t_\varepsilon)]\,\diff t_\varepsilon=\int_{t_1}^{t_2}L[q_\varepsilon(t_\varepsilon)]\,\frac{\diff t_\varepsilon}{\diff t}\,\diff t.
\end{equation*}
Then, to the first order in $\varepsilon$, one has
\begin{eqnarray*}
S(\mathcal C_\varepsilon)&=\int_{t_1}^{t_2}\Bigg(L[q(t)]+\varepsilon\mathsf X^{[1]}(L)[q(t)]\Bigg)\Bigg(1+\varepsilon\,\frac{\diff\tau(t,q(t))}{\diff t}\Bigg)\,\diff t\\
&=\int_{t_1}^{t_2}\Bigg(L[q(t)]+\varepsilon\mathsf X^{[1]}(L)[q(t)]+\varepsilon L[q(t)]\,\frac{\diff\tau(t,q(t))}{\diff t}\Bigg)\,\diff t.
\end{eqnarray*}
The first term of the integrand disappears when one forms the difference $S(\mathcal C_\varepsilon)-S(\mathcal C)$ and Formula \eref{transS} is thus obtained.

\subsection{The rearrangement \eref{relation} of the Rund-Trautman function}\label{app:I}

Using the expression \eref{firstprolongation} of the first prolongation of $\mathsf X$, one has
\begin{eqnarray*}
\mathsf X^{[1]}(L)&=\tau\,\frac{\partial L}{\partial t}+\xi^i\,\frac{\partial L}{\partial q^i}+\big(\dot\xi^i-\dot q^i\dot\tau\big)\frac{\partial L}{\partial\dot q^i}\\
&=\tau\,\frac{\partial L}{\partial t}+\big(\xi^i-\dot q^i\tau+\dot q^i\tau\big)\,\frac{\partial L}{\partial q^i}+\big(\dot\xi-\dot q^i\dot\tau-\ddot q^i\tau+\ddot q^i\tau\big)\frac{\partial L}{\partial\dot q^i}\\
&=\tau\Bigg[\frac{\partial L}{\partial t}+\dot q^i\,\frac{\partial L}{\partial q^i}+\ddot q^i\,\frac{\partial L}{\partial\dot q^i}\Bigg]+\big(\xi^i-\dot q^i\tau\big)\,\frac{\partial L}{\partial q^i}+\big(\dot\xi^i-\dot q^i\dot\tau-\ddot q^i\tau\big)\frac{\partial L}{\partial\dot q^i}\\
&=\tau\dot L+\big(\xi^i-\dot q^i\tau\big)\,\frac{\partial L}{\partial q^i}+\frac{\diff}{\diff t}\big(\xi^i-\dot q^i\tau\big)\frac{\partial L}{\partial\dot q^i}\\
&=\tau\dot L+\big(\xi^i-\dot q^i\tau\big)\,\frac{\partial L}{\partial q^i}+\frac{\diff}{\diff t}\Bigg[\big(\xi^i-\dot q^i\tau\big)\frac{\partial L}{\partial\dot q^i}\Bigg]-\big(\xi^i-\dot q^i\tau\big)\frac{\diff}{\diff t}\Bigg[\frac{\partial L}{\partial\dot q^i}\Bigg]\\
&=\tau\dot L+\big(\xi^i-\dot q^i\tau\big)\mathsf E_i(L)+\frac{\diff}{\diff t}\Bigg[\big(\xi^i-\dot q^i\tau\big)\frac{\partial L}{\partial\dot q^i}\Bigg],
\end{eqnarray*}
where use has been made of the Euler-Lagrange operators \eref{ELop}. Hence, the Rund-Trautman function \eref{RT} becomes
\begin{eqnarray*}
R=\dot B-\mathsf X^{[1]}(L)-\dot\tau L=\dot I-\big(\xi^i-\dot q^i\tau\big)\mathsf E_i(L),
\end{eqnarray*}
with
\begin{equation*}
I=B-\tau L-\big(\xi^i-\dot q^i\tau\big)\frac{\partial L}{\partial\dot q^i}=B+\tau\Bigg[\dot q^i\,\frac{\partial L}{\partial\dot q^i}-L\Bigg]-\xi^i\,\frac{\partial L}{\partial\dot q^i}\,,
\end{equation*}
and Formulas \eref{relation} and \eref{FI} are obtained.

\subsection{The gauge transformation of $\delta S(\mathcal C)$}\label{app:gauge}

Starting from
\begin{equation*}
\widetilde S(\mathcal C)=S(\mathcal C)+\Big[G(t,q(t))\Big]_{t_{1}}^{t_{2}}\quad\;\;\text{and}\quad\;\; \widetilde S(\mathcal C_\varepsilon)=S(\mathcal C_\varepsilon)+\Big[G(t_\varepsilon,q_\varepsilon(t_\varepsilon))\Big]_{t_{1\varepsilon}}^{t_{2\varepsilon}}
\end{equation*}
one has
\begin{eqnarray*}
\delta \widetilde S(\mathcal C)&=\Bigg(S(\mathcal C_\varepsilon)+\Big[G(t_\varepsilon,q_\varepsilon(t_\varepsilon))\Big]_{t_{1\varepsilon}}^{t_{2\varepsilon}}\Bigg)-\Bigg(S(\mathcal C)+\Big[G(t,q(t))\Big]_{t_{1}}^{t_{2}}\Bigg)\\
&=\Big(S(\mathcal C_\varepsilon)-S(\mathcal C)\Big)+\Bigg(\Big[G(t_\varepsilon,q_\varepsilon(t_\varepsilon))\Big]_{t_{1\varepsilon}}^{t_{2\varepsilon}}-\Big[G(t,q(t))\Big]_{t_{1}}^{t_{2}}\Bigg).
\end{eqnarray*}
The first difference is $\delta S(\mathcal C)$ while the second is equal to
\begin{equation*}
\Big(G(t_{2\varepsilon},q_\varepsilon(t_{2\varepsilon}))-G(t_{1\varepsilon},q_\varepsilon(t_{1\varepsilon}))\Big)-\Big(G(t_2,q(t_2))-G(t_1,q(t_1))\Big),
\end{equation*}
that is, to
\begin{equation*}
\Big(G(t_{2\varepsilon},q_\varepsilon(t_{2\varepsilon}))-G(t_2,q(t_2))\Big)-\Big(G(t_{1\varepsilon},q_\varepsilon(t_{1\varepsilon}))-G(t_1,q(t_1))\Big).
\end{equation*}
To the first order in $\varepsilon$, the last expression is thus equal to
\begin{equation*}
\varepsilon \mathsf X(G)(t_2,q(t_2))-\varepsilon \mathsf X(G)(t_1,q(t_1))=\varepsilon\Big[\mathsf X (G)(t,q(t))\Big]_{t_1}^{t_2}\,.
\end{equation*}
One deduces Formula \eref{vardeltaS}.

\section{Some proofs of Section \ref{sec:applications}}

\subsection{The change of variable $(t,q)\to(T,Q)$}\label{app:changevar}

Locally, along the integral curves of $\Phi$ parametrized by $\varepsilon$, the coordinates $t$ and $q$ varies according to
\begin{equation}
\frac{\diff t}{\diff\varepsilon}=\tau(t)\qquad\text{and}\qquad\frac{\diff q}{\diff\varepsilon}=\xi(t,q)=\frac12\,\dot\tau(t)q+\psi(t).\label{vartq}
\end{equation}
One seeks a new time $T$ and a new coordinate $Q$ varying as
\begin{equation}
\frac{\diff T}{\diff\varepsilon}=1\qquad\text{and}\qquad\frac{\diff Q}{\diff\varepsilon}=0.\label{varTQ}
\end{equation}
Dividing the first equality of \eref{varTQ} by the first equality of \eref{vartq}, one has
\begin{equation*}
\frac{\diff T}{\diff t}=\frac{1}{\tau(t)}\,.
\end{equation*}
This equation is automatically verified if $T$ is defined as a function of $t$ alone by
\begin{equation*}
T=\int^t\frac{\diff t'}{\tau(t')}\,.
\end{equation*}
Then, dividing the second equality of \eref{vartq} by the first, one obtains the linear differential equation
\begin{equation*}
\frac{\diff q}{\diff t}=\frac{\dot\tau(t)}{2\tau(t)}\,q+\frac{\psi(t)}{\tau(t)}\,.
\end{equation*}
It is easily integrated with the usual methods to give the relation between $q$ and $t$ along the integral curves of $\Phi$:
\begin{equation*}
q=\sqrt{\tau(t)}\Bigg(\int^t\frac{\psi(t')}{\tau(t')^{3/2}}\,\diff t'+C\Bigg)
\end{equation*}
where $C$ is a constant. Hence, the integral curves are solved as $Q=C$, with
\begin{equation*}
Q=\frac{q}{\sqrt\tau(t)}-\int^t\frac{\psi(t')}{\tau(t')^{3/2}}\,\diff t'
\end{equation*}
which thus verifies the second equality of \eref{varTQ}. The change of variables $(t,q)\to(T,Q)$ is licit since $T$ strictly increases with $t$ for any evolution and since $T$ and $Q$ are independent:
\begin{equation*}
\frac{\partial(T,Q)}{\partial(t,q)}=\left|\begin{array}{cc}
\partial_tT & \partial_qT \\
\partial_tQ & \partial_qQ
\end{array}\right|=\frac{1}{\tau(t)^{3/2}}\ne 0.
\end{equation*}
Finally, one can easily check that $\mathsf X(T)=1$ and $\mathsf X(Q)=0$, i.e.\ that $\mathsf X=\partial_T$.

\subsection{The form \eref{Valternatif} of the potential $V$}\label{app:V}

Multiplying the expression of $R_0$ in \eref{Rs} by $\tau$ yields
\begin{equation}
\tau R_0=\tau\partial_tB+\mathsf X(\tau V).\label{tauR0}
\end{equation}
Then, using the form of $B$ in \eref{taupsichi}, taking its partial derivative with respect to $t$, and expressing the result as a function of $t$ and $Q$ through
\begin{equation*}
q=\sqrt\tau\big(Q+\alpha\big)\quad\text{with}\quad\alpha(t)=\int^t\frac{\psi}{\tau^{3/2}}\,\diff t,
\end{equation*}
one obtains
\begin{equation}
\partial_tB=\frac{1}{4}\,\dddot\tau q^2+\ddot\psi q+\dot\chi=k_2(t)\,Q^2+k_1(t)\,Q+k_0(t),\label{amultiplier}
\end{equation}
with
\begin{equation*}
k_2(t)=\frac14\,\dddot\tau\tau\;\,,\quad k_1(t)=\ddot\psi\sqrt\tau+\frac12\,\alpha\dddot\tau\tau\;\,,\quad k_0(t)=\alpha\ddot\psi\sqrt\tau+\frac14\,\alpha^2\dddot\tau\tau+\dot\chi.
\end{equation*}
One remarks that the variable $\rho(t)=\sqrt \tau$ is such that
\begin{equation}
\frac12\,\dddot\tau\tau=\frac{\diff}{\diff t}\Big(\rho^3\ddot\rho\Big).\label{dddtau}
\end{equation}
Hence
\begin{equation*}
k_2(t)=\frac{\diff K_2(t)}{\diff t}\quad\text{with}\quad K_2(t)=\frac12\,\rho^3\ddot\rho.
\end{equation*}
Applying two times the Leibniz rule on the product $\psi\rho$ and using the fact that $\psi=\tau^{3/2}\dot\alpha=\rho^3\dot\alpha$, one obtains
\begin{equation}
\ddot\psi\sqrt\tau=\ddot\psi\rho=\frac{\diff}{\diff t}\Big(\dot\psi\rho-\psi\dot\rho\Big)+\psi\ddot\rho=\frac{\diff}{\diff t}\Bigg[\rho^2\,\frac{\diff(\rho^2\dot\alpha)}{\diff t}\Bigg]+\dot\alpha\rho^3\ddot\rho\label{ddpsi}
\end{equation}
and thus
\begin{equation*}
k_1(t)=\frac{\diff}{\diff t}\Bigg[\rho^2\,\frac{\diff(\rho^2\dot\alpha)}{\diff t}\Bigg]+\dot\alpha\rho^3\ddot\rho+\alpha\, \frac{\diff}{\diff t}\Big(\rho^3\ddot\rho\Big),
\end{equation*}
that is,
\begin{equation*}
k_1(t)=\frac{\diff K_1(t)}{\diff t}\quad\text{with}\quad K_1(t)=\rho^2\Big(\rho^2\ddot\alpha+2\rho\dot\rho\dot\alpha+\rho\ddot\rho\alpha\Big).
\end{equation*}
Then, using \eref{dddtau} and \eref{ddpsi}, one has
\begin{eqnarray*}
k_0(t)&=\alpha\,\frac{\diff}{\diff t}\Bigg[\rho^2\,\frac{\diff(\rho^2\dot\alpha)}{\diff t}\Bigg] +\alpha\dot\alpha\rho^3\ddot\rho+\frac12\,\alpha^2\,\frac{\diff}{\diff t}\Big(\rho^3\ddot\rho\Big)+\dot\chi\\
&=\frac{\diff}{\diff t}\Bigg[\alpha\rho^2\,\frac{\diff(\rho^2\dot\alpha)}{\diff t}\Bigg]-\dot\alpha\rho^2\,\frac{\diff(\rho^2\dot\alpha)}{\diff t} +\frac{1}{2}\frac{\diff(\alpha^2)}{\diff t}\,\rho^3\ddot\rho+\frac12\,\alpha^2\,\frac{\diff}{\diff t}\Big(\rho^3\ddot\rho\Big)+\dot\chi.
\end{eqnarray*}
The second term is clearly the $t$-derivative of $-(\rho^2\dot\alpha)^2/2$ while the sum of the two next terms is the $t$-derivative of $\alpha^2\rho^3\ddot\rho/2$. Hence, one has
\begin{equation*}
k_0(t)=\frac{\diff K_0(t)}{\diff t}\quad\text{with}\quad K_0(t)=\rho^2\bigg(\rho^2\alpha\ddot\alpha+2\rho\dot\rho\alpha\dot\alpha+\frac12\,\rho\ddot\rho\alpha^2-\beta\bigg)
\end{equation*}
where was introduced the quantity
\begin{equation*}
\beta(t)=\frac{\psi^2}{2\tau^2}-\frac{\chi}{\tau}=\frac12\,\rho^2\dot\alpha^2-\frac{\chi}{\rho^2}\,.
\end{equation*}
Consequently, the multiplication of \eref{amultiplier} by $\tau$ gives
\begin{eqnarray*}
\tau\partial_tB&=\tau\,\frac{\diff K_2}{\diff t}\,Q^2+\tau\,\frac{\diff K_1}{\diff t}\,Q+\tau\,\frac{\diff K_0}{\diff t}
=\frac{\diff K_2}{\diff T}\,Q^2+\frac{\diff K_1}{\diff T}\,Q+\frac{\diff K_0}{\diff T}\\
&=\frac{\partial}{\partial T}\Big(K_2Q^2+K_1Q+K_0\Big)_Q
\end{eqnarray*}
and Equation \eref{tauR0} becomes
\begin{equation}
\rho^2R_0=\frac{\partial}{\partial T}\Big(K_2Q^2+K_1Q+K_0+\rho^2V\Big)
\end{equation}
since $\mathsf X=\partial_T$ in the adapted system. Therefore, taking an integral with respect to $T$ (with $Q$ kept constant), there is a function $W(Q)$ such that
\begin{equation}
\int^T\hspace{-2mm}\rho^2 R_0\,\diff T=K_2Q^2+K_1Q+K_0+\rho^2V-W(Q).\label{KV}
\end{equation}
It still remains to isolate $V$ and reexpress $Q$ in terms of the old coordinates to obtain Formula \eref{Valternatif}.

\subsection{The gauge term \eref{Lambda}}\label{app:Lambda}

We seek a solution $G$ to the gauge condition $B+\mathsf X(G)=0$ with
\begin{equation*}
B=\frac14\,\ddot\tau q^2+\dot\psi q+\chi.
\end{equation*}
Expressing $q$ as a function of $t$ and $Q$, one has
\begin{equation}
B=\tau\Big(\ell_2(t)Q^2+\ell_1(t)Q+\ell_0(t)\Big),\label{BQ}
\end{equation}
with
\begin{equation*}
\ell_2(t)=\frac14\,\ddot\tau\quad,\quad\ell_1(t)=\frac12\,\alpha\ddot\tau+\frac{\dot\psi}{\sqrt\tau}\quad,\quad\ell_0(t)=\frac14\,\alpha^2\ddot\tau+\alpha\,\frac{\dot\psi}{\sqrt\tau}+\frac{\chi}{\tau}\,.
\end{equation*}
One has obviously
\begin{equation*}
\ell_2(t)=\frac{\diff L_2(t)}{\diff t}\quad\text{with}\quad L_2(t)=\frac14\,\dot\tau=\frac12\,\rho\dot\rho.
\end{equation*}
Then, using the Leibniz rule as above, one has
\begin{equation*}
\ell_1(t)=\frac{\diff}{\diff t}\Bigg(\frac12\,\alpha\dot\tau\bigg)-\frac12\,\dot\alpha\dot\tau+\frac{\dot\psi}{\sqrt\tau}=\frac{\diff}{\diff t}\Bigg(\frac12\,\alpha\dot\tau\bigg)-\frac{\psi\dot\tau}{2\tau^{3/2}}+\frac{\dot\psi}{\sqrt\tau}\,,
\end{equation*}
that is,
\begin{equation*}
\ell_1(t)=\frac{\diff L_1(t)}{\diff t}\quad\text{with}\quad L_1(t)=\frac12\,\alpha\dot\tau+\frac{\psi}{\sqrt\tau}=\alpha\rho\dot\rho+\dot\alpha\rho^2.
\end{equation*}
As for $\ell_0(t)$, one has
\begin{eqnarray*}
\ell_0(t)&=\frac{\diff}{\diff t}\Bigg(\frac14\,\alpha^2\dot\tau\Bigg)-\frac12\,\dot\alpha\alpha\dot\tau+\alpha\,\frac{\dot\psi}{\sqrt\tau}+\frac{\chi}{\tau}\\
&=\frac{\diff}{\diff t}\Bigg(\frac14\,\alpha^2\dot\tau\Bigg)+\alpha\Bigg(-\frac{\psi\dot\tau}{2\tau^{3/2}}+\frac{\dot\psi}{\sqrt\tau}\Bigg)+\frac{\chi}{\tau}\\
&=\frac{\diff}{\diff t}\Bigg(\frac14\,\alpha^2\dot\tau\Bigg)+\alpha\,\frac{\diff}{\diff t}\Bigg(\frac{\psi}{\sqrt\tau}\Bigg)+\frac{\chi}{\tau}\\
&=\frac{\diff}{\diff t}\Bigg(\frac14\,\alpha^2\dot\tau+\alpha\,\frac{\psi}{\sqrt\tau}\Bigg)-\dot\alpha\,\frac{\psi}{\sqrt\tau}+\frac{\chi}{\tau}\\
&=\frac{\diff}{\diff t}\Bigg(\frac12\,\rho\dot\rho\alpha^2+\rho^2\alpha\dot\alpha\Bigg)-\frac12\,\rho^2\dot\alpha^2-\beta,
\end{eqnarray*}
that is,
\begin{eqnarray*}
\ell_0(t)=\frac{\diff L_0(t)}{\diff t}\quad\text{with}\quad L_0(t)=\frac12\,\rho\dot\rho\alpha^2+\rho^2\alpha\dot\alpha-\int^t\Bigg(\frac12\,\rho^2\dot\alpha^2+\beta\Bigg)\diff t.
\end{eqnarray*}
Therefore,
\begin{eqnarray*}
B&=\tau\,\frac{\diff L_2}{\diff t}\,Q^2+\tau\,\frac{\diff L_1}{\diff t}\,Q+\tau\,\frac{\diff L_0}{\diff t}=\frac{\diff L_2}{\diff T}\,Q^2+\frac{\diff L_1}{\diff T}\,Q+\frac{\diff L_0}{\diff t}\\
&=\frac{\partial}{\partial T}\Big(L_2Q^2+L_1Q+L_0\Big)_Q,
\end{eqnarray*}
and the gauge condition $B+\mathsf X(G)=0$ reduces to
\begin{equation*}
\frac{\partial}{\partial T}\Big(L_2Q^2+L_1Q+L_0+G\Big)_Q=0.
\end{equation*}
It suffices to choose
\begin{equation}
G=-L_2Q^2-L_1Q-L_0\,.\label{GQ}
\end{equation}
Replacing $Q$ with its expression as a function of $t$ and $q$, one obtains Formula \eref{Lambda}.

\subsection{The adapted Lagrangian \eref{newlag}}\label{app:L}

The adapted Lagrangian $\overline L$ is such that $\overline L\diff T=L\diff t+\diff G$, i.e.\
\begin{equation*}
\overline L=L\,\frac{\diff t}{\diff\tau}+\frac{\diff G}{\diff T}=L\,\rho^2+\frac{\partial G}{\partial T}+\mathring Q\,\frac{\partial G}{\partial Q}=\rho^2\Bigg(\frac12\,\dot q^2-V\Bigg)-B+\mathring Q\,\frac{\partial G}{\partial Q}
\end{equation*}
where the empty bullet symbolizes the total $T$-derivative. Then, using
\begin{equation*}
\dot q=\frac{\diff q}{\diff t}=\frac{\diff}{\diff t}\Big[\rho(Q+\alpha)\Big]=\rho\,\dot Q+\dot\rho Q+\frac{\diff(\rho\alpha)}{\diff t}=\frac{\mathring Q}{\rho}+\dot\rho Q+(\dot\rho\alpha+\rho\dot\alpha)
\end{equation*}
together with the expressions \eref{KV}, \eref{BQ} and \eref{GQ} of $V$, $B$ and $G$ respectively, one obtains
\begin{eqnarray*}
\overline L=&\frac12\,\mathring Q^2+M_4(t)\mathring QQ+M_3(t)\mathring Q+M_2(t)Q^2+M_1(t)Q+M_0(t)\\
&-W(Q)-\int^T\hspace{-2mm}\rho^2R_0\,\diff T
\end{eqnarray*}
with
\begin{equation*}
\begin{array}{ll}
M_4(t)=\rho\dot\rho-2L_2\,, & M_1(t)=\rho^2\dot\rho(\dot\rho\alpha+\rho\dot\alpha)+K_1-\rho^2\ell_1\,,\vphantom{\bigg|}\\
M_3(t)=\rho(\dot\rho\alpha+\rho\dot\alpha)-L_1\,, & M_0(t)=\displaystyle\frac12\,\rho^2(\dot\rho\alpha+\rho\dot\alpha)^2+K_0-\rho^2\ell_0\,.\\
M_2(t)=\displaystyle\frac12\,\rho^2\dot\rho^2+K_2-\rho^2\ell_2\,,\;\;\;\;\quad
\end{array}
\end{equation*}
But one verifies with the definitions of the various quantities introduced in \ref{app:V} and \ref{app:Lambda} that the $M_i(t)$s identically vanish. The adapted Lagrangian \eref{newlag} is thus obtained.

\bibliography{adiabaticN}

\end{document}